\documentstyle[epsf]{fbsart11}
\newcommand{\CA}{{\cal A}}

\newcommand{\CD}{{\cal D}}
\newcommand{\CM}{{\cal M}}
\newcommand{\CB}{{\cal B}}
\newcommand{\CC}{{\cal C}}
\newcommand{\CF}{{\cal F}}
\newcommand{\CO}{{\cal O}}
\begin{document}
\def\bp{{\mbox{\boldmath$p$}}}
\def\bn{{\mbox{\boldmath$n$}}}
\def\ba{{\mbox{\boldmath$a$}}}
\def\bxi{{\mbox{\boldmath$\xi$}}}
\def\bsigma{{\mbox{\boldmath$\sigma$}}}
\def\thesection{\Roman{section}}

\title{
Pion Exchange Effects in Elastic Backward \\
Proton-Deuteron Scattering
}

\author{
L.P. Kaptari\instnr{1,2}
B. K\"ampfer\instnr{1}
S.M. Dorkin\instnr{3}
S.S. Semikh\instnr{2}
}

\instlist{
Research Center Rossendorf, Institute for Nuclear and Hadron Physics,
PF 510119, 01314 Dresden, Germany
\and
Bogoliubov Laboratory of Theoretical Physics, JINR Dubna,
P.O. Box 79, Moscow, Russia
\and
Far-Eastern State University, Vladivostok, Russia
}
\runningauthor{L.P. Kaptari et al.}
\runningtitle{Pion exchange effects in elastic backward $pD$ scattering}

\maketitle

\begin{abstract}
The elastic backward proton-deuteron scattering is analyzed
within a covariant approach based on the Bethe-Salpeter
equation with realistic meson-exchange interaction.
Contributions of the one-nucleon and one-pion
exchange mechanisms to the cross section and polarization
observables are investigated in explicit form.
Results of numerical calculations for the cross section,
tensor analyzing power and spin transfers are presented.
The one-pion exchange contribution is essential for
describing  the spin averaged cross section, while
in polarization observables it is found to be less important.
\end{abstract}


\section{Introduction}

Presently a wide-spread program of investigations of the structure of
the lightest nuclei is under consideration.
There are new proposals to study the polarization
characteristics of the deuteron using both hadron
\cite{proposal,cosy,preliminar}
and electro-magnetic probes (cf.\ refs.~\cite{arenh}).
The future experiments  are very interesting since
a complete reconstruction of the
amplitude of the process~\cite{proposal,arenh,rekalo} and
an overall investigation of the nuclear momentum distribution
\cite{cosy,experiment} seems achievable.

Among the simplest reactions with hadron probes are
processes of forward or backward scattering of protons
off the deuteron. The intensive experimental study of these
reactions has started a decade ago at Dubna
and Saclay (see, for instance, refs.
\cite{experiment,dpelastic,punj95,belost,aleshin})
and is also planed to be continued in the nearest future at COSY ~\cite{cosy}.

The measured momentum of the fragments
is directly related to  the argument of the deuteron
wave function in the momentum space. In this way one
 hopes that
a direct experimental investigation of the momentum distribution
in the deuteron in a large interval of the internal momenta is possible.
By using  polarized particles one may investigate as well
different aspects of spin-orbital interaction in the deuteron
and can obtain important information about the role of
``non-traditional'' degrees of freedom (like $\Delta$ isobars,
$N\bar N$ excitation and so on) in the deuteron wave function.

Nowadays a renewed interest receives the
elastic backward $pD$ scattering with  polarized protons and deuterons.
A distinguished peculiarity of this process is that
in the impulse approximation the cross section is
proportional to the fourth power of the deuteron wave
function (contrary to the break-up and quasi-elastic
reactions which are  proportional to the
second power of the wave function).
This makes the processes of elastic scattering much more
sensitive to the theoretically assumed mechanisms and
the deuteron wave function.
Another peculiarity  of elastic backward or forward  $pD$
reactions is that the amplitude of the process is
determined by only four complex helicity amplitudes,
and a set-up of experiments for a complete reconstruction
of these amplitudes is rather possible. For this
it is sufficient to measure 10 independent observables
as proposed in refs.~\cite{rekalo,ladygin}.

First measurements of polarization observables, such as  the
tensor analyzing power $T_{20}$ and deuteron-proton polarization transfer
$\kappa_{D\to p}$, have been performed at Dubna~\cite{dpelastic,azh97}
and Saclay \cite{punj95,belost}.
Theoretically the elastic $pD$ scattering has been studied by several authors
\cite{ourphysrev,santos,kislind,keisterTj,wilkin,nakamura,kolybasov}.
It has been shown that
the cross section can not be satisfactorily described within
the  impulse approximation and that other
mechanisms, such as meson exchange or triangle
diagrams \cite{wilkin,nakamura,kolybasov}, may become important.
It is expected that the role of virtual
meson production becomes important
at moderate energies of the incoming proton,
$p\sim 1.2 \, - \, 1.3$ GeV/c (or equivalently, at momenta of the
outgoing proton    $P_{lab}\sim 0.31 \,- \,0.33$ GeV/c),
corresponding to the range of
excitations of $\Delta$ isobars during the interaction.
Namely at these energies the experimental cross section
exhibits a relatively broad bump (cf. \cite{dpelasexp}), which
may be considered as  an evidence of the $\Delta$ excitations
in the process of elastic backward $pD$ scattering. Consequently,
an investigation of the contribution of such diagrams to the
unpolarized cross section and to the
polarization observables, which are more sensitive to
the reaction mechanism, is of interest.

We focus our attention here on a  study of the contribution
of virtual meson production in the elastic
$pD$ amplitude within the Bethe-Salpeter (BS)
approach by using a numerical solution of the BS equation
obtained with a realistic one-boson exchange
interaction \cite{solution,parametrization}.
For the triangle diagrams we consider
only the positive energy waves (the largest ones), because the
negative $P$ waves are expected to provide a  negligible contribution
to the total amplitude. Within the present approach the effects of
relativistic Fermi motion, Lorenz boosts and spin transformations of
the relevant amplitudes are treated in a fully covariant way.
The total amplitude of the process is presented as a sum
of the amplitudes of the one-nucleon exchange
mechanism (investigated in details in a  previous paper \cite{ourphysrev})
and the one-meson exchange diagrams. It is explicitly
shown that the total amplitude is time-reversal invariant.

Our paper is organized as follows. In section II we present the
kinematics and notation. Section III contains the relevant
formulae of both the one-nucleon exchange process and, in more
detail, the triangle diagrams.
Results (such as numerical calculations of the cross section,
tensor analyzing power $T_{20}$, polarization  transfer from deuteron to
proton $\kappa_{D\to p}$ and from proton
to proton $\kappa_{p\to p}$) are discussed in section IV and
summarized in section V. Some important technical details are presented in
the appendices.

\section{ Kinematics and notation}

We consider the  elastic backward $pD$ scattering of the type
\begin{equation}
 p\,\,+\,\,D\,\,=\,\, p'(\theta_{cm}=180^0)\,\,+D' .
\label{reaction}
\end{equation}
The differential cross section of the reaction (\ref{reaction})
in the center of mass of the colliding particles reads
\begin{equation}
\frac{d\sigma}{d\Omega}\,=\,\frac{1}{64\pi^2S}\,
\frac{1}{6}\sum_{spin}\,|\CM_{s,M}^{s',M'}\,|^2,
\label{cross}
\end{equation}
where $S$ is the Mandelstam variable
denoting the total energy in the center of mass,
and $\CM$ is the invariant amplitude of the process. In the
case of backward scattering the cross section Eq.~(\ref{cross}) depends
only on one  kinematical variable, which  is usually  chosen to be
$S$. Other variables can be expressed via $S$ by using the
energy conservation law; for instance, the Mandelstam variable $u$
is $u=(M_D^2-m^2)^2/S$, the center of mass momentum
reads $\bp\, ^{2}=-t/4$ etc.
Here $M_D$ and $m$ stand for the deuteron and nucleon masses,
respectively.
In the laboratory  system
we define the relevant kinematical variables as follows
\begin{equation}
D=(M_D,{\bf 0}),\,\,
q=\frac{1}{2}(p_1-p_2)=(q_0,{\bf q}),\,\,\,p'=(E_{lab},\bp'), \, \,\,
|\bp'|=P_{lab},
\label{moms}
\end{equation}
and the components of the polarization  vector of the deuteron
with the polarization index $M=\pm 1,\, 0$ are fixed by
\begin{equation}
\begin{array}{ccc}
\bxi_{+ 1} =-\displaystyle\frac{1}{\sqrt{2}}\,
\left ( \begin{array}{c} 1 \\ \i \\ 0 \end{array} \right ),
&
\quad \bxi_{- 1}=\displaystyle\frac{1}{\sqrt{2}}
\left ( \begin{array}{c}1\\ -\i\\0 \end{array} \right ),
&
\quad \bxi_{0}=
\left ( \begin{array}{c} 0\\ 0\\1 \end{array} \right ).
\label{xilab}
\end{array}
\end{equation}
The Dirac spinors, normalized  as $\bar{u}(p) u(p)=2m$
and $\bar{v}(p) v(p)=-2m$, read
\begin{eqnarray}
u(\bp,s)=\sqrt{m+\epsilon}
\left ( \begin{array}{c} \chi_s \\ \frac{\bsigma \bp}{m+\epsilon}\chi_s
\end{array}  \right), \quad \quad
v(\bp\,,s)=\sqrt{m+\epsilon}\left( \begin{array}{c}
\frac{\bsigma \bp}{m+\epsilon}\widetilde\chi_{s}\\
 \widetilde\chi_{s}
\end{array}  \right),
\label{spinors}
\end{eqnarray}
where $\widetilde\chi_s\equiv - \i \sigma_y\chi_s$,  and
$\chi_s$ denotes the usual two-dimensional Pauli spinor .

The  general properties
of  the   amplitude  $\CM$ for the elastic scattering of
the spin type $1/2\,+\,1\,=\,1/2\,+\,1$
have been  studied in detail and are well  known
(see, for instance, refs.~\cite{rekalo,ladygin});
here we recall only   the most important characteristics of $\CM$.
In principle, the process of the elastic $pD$ scattering
is  determined by 12 independent partial amplitudes.
However, in case of forward or backward scattering, because of
the conservation of the total spin projection,
only four amplitudes remain independent and these four
amplitudes determine all the possible polarization observables
of the process.
There are many possible choices of representing these
four amplitudes. In order to emphasize explicitly the
transition between initial and final states
with fixed spin projections it is convenient to represent
$\CM$ in the center of mass in the
two-dimensional spin space of  the
proton spinors  and three-dimensional space of the
deuteron spin characteristics.
In such a representation of the amplitude it is possible,
by  using  Eqs.~(\ref{moms}) - (\ref{spinors}),
to express all the partial spin amplitudes
via the corresponding quantities
evaluated in the deuteron's rest frame.
Then using the same notation as
adopted in  refs.~\cite{rekalo,ladygin,ourphysrev},
the total amplitude is written in the form
\begin{equation}
\CM_{s,M}^{s',M'} =  \chi_{s'}^+\, \CF_{M,M'} \,\chi_s ,
\label{ampF}
\end{equation}
with
\begin{eqnarray}
\CF_{M,M'} & = &
{\cal A}\,(\bxi_M\bxi^+_{M'})+
{\cal B}\,(\bn\bxi_M)(\bn\bxi^+_{M'})
+ \i {\cal C}\,(\bsigma\cdot [\bxi_M\times\bxi^+_{M'}])\nonumber\\[2mm]
& + &
\i \, {\cal D}\,(\bsigma\bn)(\bn\cdot [\bxi_M\times\bxi^+_{M'}]) ,
\label{amplit}
\end{eqnarray}
where $\bn$ is a unit vector parallel to the beam direction, and
${\cal A,B,C}, {\cal D}$ are the partial amplitudes of the $pD$ elastic
scattering depending on the initial energy.
The cross section (\ref{cross}) for unpolarized particles is determined
by $\mbox{Tr} \left ( \CF^+\,\CF \right )$.
The four scalar amplitudes ${\cal A},{\cal B},{\cal C}$, ${\cal D}$
are related to the partial spin amplitudes $\CM_{s,M}^{s',M'}$ via
\begin{eqnarray}
&&
\CM_{\frac{1}{2}1}^{\frac{1}{2}1}=\CA+\CC+\CD ,\quad
\CM_{\frac{1}{2}0}^{\frac{1}{2}0}=\CA+\CB ,\nonumber\\[2mm]
&&
\CM_{\frac{1}{2}-1}^{\frac{1}{2}-1}=\CA - \CC -\CD,\quad
\CM_{\frac{1}{2}0}^{-\frac{1}{2}1}=\sqrt{2}\CC .\label{spinampl}
\end{eqnarray}

The differential cross section and polarization
observables under consideration read \cite{ladygin}
\begin{eqnarray}
\frac{d\sigma}{d\Omega} & = &
\frac{1}{64\pi^2S}\,
\frac{1}{6} \mbox{Tr} \left (\CF\CF^+\right ) ,
\label{cross1}\\[2mm]
T_{20} & = &
-\frac{1}{\sqrt{2}}
\frac{4\left(
2 \Re \left [ \CA\CB^* \right ] +|\CB|^2 -
2 \Re \left [\CC\CD^*\right ]-|\CD|^2\right )}
{\mbox{Tr} \left (\CF\CF^+\right )},\label{t20}\\[2mm]
\kappa_{D\to p} & = &
\frac{6\left(
\Re \left [ (2\CA^*+\CB^*+\CD^*)\CC \right ] +|\CC|^2 \right )}
{\mbox{Tr} \left (\CF\CF^+\right )},\label{dtop}\\[2mm]
\kappa_{p\to p} & = &
\frac{2 \left( 3|\CA|^2 +2 \Re  (\CA\CB^*)
+ |\CB|^2 -2|\CC|^2 -4 \Re( \CC\CD^*)-2|\CD|^2\right )}
{ \mbox{Tr} \left (\CF\CF^+\right )},\label{ptop}
\end{eqnarray}
with
\begin{equation}
\mbox{Tr} \left (\CF\CF^+\right ) = 2(3|\CA|^2 + 2 \Re
\left [ \CA\CB^*\right ] +|\CB|^2+6|\CC|^2+
4 \Re \left [ \CC\CD^*\right ] +2|\CD|^2).
\label{trace}
\end{equation}

\section{ Basic formulae}

In what follows we investigate the contribution
of the relativistic one-nucleon exchange  and
one-meson exchange diagrams to the process (\ref{reaction})
as depicted in Figs.~1a - 1c. Correspondingly, the amplitude
reads
\begin{equation}
\CM_{s,M}^{s',M'} = {\cal T}_{s,M}^{s',M'}+ T_{s,M}^{s',M'}+\widetilde T_{s,M}^{s',M'},
\label{three}
\end{equation}
where ${\cal T}_{s,M}^{s',M'}$ stands for the one-nucleon exchange amplitude, and
$T_{s,M}^{s',M'}$ and $\widetilde T_{s,M}^{s',M'}$ denote the contributions
of the triangle diagrams in Figs.~1b and 1c.

\subsection{The one-nucleon exchange diagram}

The contribution of the one-nucleon exchange mechanism
in $pD$ reactions has been investigated within the BS
formalism in detail elsewhere (cf.\
refs.~\cite{ourphysrev,kislind,keisterTj,break}). Therefore,  we
briefly recall only  the main
results for the one-nucleon exchange diagram.

Using the kinematics shown in Fig.~1 the  one-nucleon
exchange contribution to the elastic amplitude within the
BS formalism is
\begin{eqnarray}
{\cal T}_{s,M}^{s',M'} = \bar u(\bp',s') \, \Gamma_M (p_2,p') \,
\tilde{S_2} \, \bar{\Gamma}_{M'} (p_2,p) \, u(\bp,s),
\label{gn}
\end{eqnarray}
where $\Gamma(p_1,p_2)$ denotes the BS vertex function
of the deuteron, $\tilde S_2 = (\hat p_2+m)^{-1}$ is the
modified propagator of the exchanged (second) particle (see Fig.~1a).
By making use of Eqs.~(\ref{moms}) - (\ref{spinors}) the covariant
amplitude (\ref{gn}) may be expressed in the form (\ref{amplit}).
The result (only for positive BS waves) is \cite{ourphysrev}:
\begin{eqnarray}
\frac{d\sigma_0}{d\Omega} & = &
\frac{12m^2}{S}\,\left( \Psi_S^2(P_{Lab})
+ \Psi^2_D(P_{Lab})\right)^2\,P_{Lab}^4,
\label{crospositive}
\\[2mm]
{\cal A}_0 & = &
16\pi m P_{Lab}^2\,\left(
\Psi_S(P_{Lab})-\frac{\Psi_D(P_{Lab})}{\sqrt{2}}\right) ^2 ,
\label{aBS}\\
{\cal B}_0 & = &
16\pi m P_{Lab}^2\
\frac{3}{2}\Psi_D(P_{Lab})\,\left( 2\sqrt{2}\Psi_S(P_{Lab})
+\Psi_D(P_{Lab})\right ),
\label{bBS}\\
{\cal C}_0 & = &
16\pi mP_{Lab}^2\,\left( \Psi_S(P_{Lab})
-\frac{\Psi_D(P_{Lab})}{\sqrt{2}}\right)
\left(\Psi_S(P_{Lab})+\sqrt{2}\Psi_D(P_{Lab})\right),
\label{cBS}\\
{\cal D}_0 & = &
-\,16\pi mP_{Lab}^2\,
\frac{3}{\sqrt{2}}\Psi_D(P_{Lab})
\left( \Psi_S(P_{Lab})-\frac{\Psi_D(P_{Lab})}{\sqrt{2}}\right),
\label{dBS}
\end{eqnarray}
where  we employ the notion of BS wave functions with
\cite{keisterTj,gross,quad}
\begin{equation}
\Psi_S(|{\bf P}_{Lab}|) = {\cal N}
\frac{G_{S^{++}}(p_0,|{\bf P}_{Lab}|)}{2E_{lab}-M_D},
\quad
\Psi_D(|{\bf P}_{Lab}|) = {\cal N}\frac{G_{D^{++}}(p_0,|{\bf P}_{Lab}|)}{2E_{lab}-M_D},
\label{positive}
\end{equation}
with $p_0=M_D-2 E_{Lab}$,  ${\cal N} = 1/4\pi\sqrt{2M_D}$, and
$G_{S,D}(p_0,|{\bf P}_{Lab}|)$ as  the partial vertices with  positive
$\rho$ spins.

Note, that within the one-nucleon exchange approximation
the BS formalism with only positive-energy waves provides  exactly the same
form of polarization observables
as in the non-relativistic impulse approximation~\cite{rekalo,ladygin},
whereas an account of the negative-energy  $P$ waves
leads to  more complicate relations among
the polarization observables and the deuteron wave function \cite{ourphysrev}.

\subsection{The triangle diagram}

Now we proceed with an investigation of the triangle diagram depicted in
Fig.~1b. The corresponding amplitude, in case when the exchanged
particles are a proton and a  $\pi^+$, has the form
\begin{equation}
T_{s,M}^{s',M'} =  \i \int\frac{d^4 q}{(2\pi)^4} \,
\bar u(\bp',s') \,
\frac{\sqrt{2} g_{\pi NN}
\gamma_5}{\pi^2 - \mu^2 +\i \varepsilon} \,
\Phi_M(p_1,p_2) \,
\hat A^{M'}_{p_2+p\to\pi^++ D'} \,
u(\bp, s),
\label{triangl1}
\end{equation}
where
$g_{\pi NN}$ is the $\pi NN$ coupling constant,
$\mu$ is the pion mass, $ \Phi_M(p_1,p_2)$  the BS amplitude,
and the operator
$\hat A^{M'}_{p_2+p\to\pi^++ D'}$ describes the off-mass shell process
$ {p_2+p\to\pi^++ D'}$. Besides the amplitude (\ref{triangl1}),
there is another
diagram with neutron and $\pi^-$ exchange, which is related to
(\ref{triangl1}) by an isospin factor of $1/2$.

For the sake of consistency, the operator $\hat A $ which is a
$4\times 4$ matrix in the spinor space should be computed within
the same framework of the effective meson-nucleon theory as  the BS
equation is solved. However, this is a rather cumbersome and 
ambitiouse problem.
To avoid uncertainties connected with  calculations of  $\hat A $, in
this paper we express it via the amplitude $f_{s_2,s}^{M'}$
of the subprocess $p+p\to \pi^+ +D$
with real  on-mass shell particles by
\begin{equation}
f_{s_2,s}^{M'} = \langle \pi, D'|\hat A |p_2,p\rangle =
\bar v(\bp_2,s_2) \,
\hat A^{M'} \,
u(\bp,s).
\label{fpiD}
\end{equation}
The BS  amplitude $ \Phi_M(p_1,p_2)$ is related to the deuteron BS vertex
\begin{equation}
\Phi_M(p_1,p_2) =
\frac{\hat p_1+m }{\left (p_1^2-m^2 + \i \varepsilon \right )} \,
\Gamma_{M} (p_1,p_2) \,
\frac{\hat p_2-m}{\left ( p_2^2-m^2 + \i \varepsilon\right )}.
\label{vert}
\end{equation}
It is seen, at first sight, from Eqs.~(\ref{triangl1}) and (\ref{vert}) that
there are six poles in the integrand (\ref{triangl1}): three located in the
upper half-plane of $q_0$, the other three ones in the lower half-plane.
Actually, the two  nucleon propagators in (\ref{vert})
contain only two poles, as it may be shown,
for instance, by performing  an explicit partial decomposition
of the amplitude $ \Phi_M(p_1,p_2)$ in the  spin-angular basis \cite{quad},
hence the integrand (\ref{triangl1})  contains two poles in the
upper half-plane and  two poles in the lower one.

Inserting  Eqs.~(\ref{fpiD}) and (\ref{xilab}) - (\ref{spinors})
into Eq.~(\ref{triangl1})
and keeping in
the amplitude $ \Phi_M(p_1,p_2)$ only  partial waves with positive relative energy
 (the explicit form of  $ \Phi_M(p_1,p_2)$ in terms of positive waves
is given in  Appendix \ref{sec:app1})
the contribution of the triangle diagram reads
\begin{eqnarray}
T_{s,M}^{s',M'} & = &
\frac{\i g_{\pi NN}}{\sqrt{2\pi}} \sum_{s_2}
\int\frac{d^4 q}{(2\pi)^4}
\frac{1} {\pi^2 - \mu^2 + \i \varepsilon}
\frac{1}{E}\sqrt{\displaystyle\frac{E_{lab}+m}{E_q+m}}
f^{M'}_{s_2,s}(p,p_2;\pi ,D')
\label{triang3}\\[2mm]
& \times &
\left[
\frac{ G_{S^{++}} (q_0,|{\bf q}|) - G_{D^{++}} (q_0,|{\bf q}|)/\sqrt{2} }
{(M_D-2E_q)^2-4q_0^2 + \i \varepsilon}
\chi^+_{s'}
\left( (\bsigma {\bf \cal K})(\bsigma\bxi_M) \right)
\widetilde \chi_{s_2}
\right. \nonumber\\[2mm]
+ &&
\left.
\frac{3\,  G_{D^{++}} (q_0,|{\bf q}|) }
{(M_D-2E_q)^2-4q_0^2 + \i \varepsilon} (\bxi_M {\bf q})
\chi^+_{s'}
\left( 1 -\frac{E_q+m}{E_{lab}+m}
\frac{(\bsigma {\bf q} )(\bsigma \bp)}{{\bf q}^2} \right)
\widetilde \chi_{s_2}
\right] ,
\nonumber
\end{eqnarray}
where ${\bf \cal K}={\bf q}-(E_q+m)/(E_{lab}+m)\bp'\,$,
$\, E_q=\sqrt{{\bf q}^2+m^2}$ and $E_{lab}=\sqrt{P_{lab}^2+m^2}$.

Equation~(\ref{triang3}) may be simplified by
observing that the amplitude $T_{s,M}^{s',M'}$
is  a function of  external momenta of the process (\ref{reaction}),
for which only one independent three-vector may be defined, for instance
the vector ${\bf n}$ from (\ref{amplit}). Then one obtains
\begin{eqnarray}
&&
T_{s,M}^{s',M'}= \sum_{s_2}
\chi^+_{s'} \left [ a_{s_2,s}^{M'}\, (\bsigma {\bf n})(\bsigma\bxi_M)
+ d_{s_2,s}^{M'} (\bf n\bxi_M)\right ] \widetilde\chi_{s_2}, \label{simple}
\end{eqnarray}
where we introduced two scalar functions as follows
\begin{eqnarray}
a_{s_2,s}^{M'}& = &
-\frac{\i g_{\pi NN}}{\sqrt{4\pi}}
\int\frac{d^4 q}{(2\pi)^4}\frac{1} {\pi^2 - \mu^2 + \i \varepsilon}
\frac{2}{E_q} \sqrt{\displaystyle\frac{E_{lab}+m}{E_q+m}}
\frac{1} {\pi^2 - \mu^2 + \i \varepsilon}
\label{triang4}\\
& \times &
\frac{ G_{S^{++}}
(q_0,|{\bf q}|)
-G_{D^{++}}(q_0,|{\bf q}|) \frac{1}{\sqrt{2}}
}{(M_D-2E_q)^2-4q_0^2 + \i \varepsilon}
\left( |{\bf q}|\cos\theta +\frac{E_q+m}{E_{lab}+m} |\bp'|\right)
\nonumber\\[2mm]
& \times &
f^{M'}_{s_2,s}(p,p_2;\pi ,D'),
\nonumber\\[2mm]
d_{s_2,s}^{M'}& = &
-\frac{\i 3g_{\pi NN}}{\sqrt{8\pi}}
\int\frac{d^4 q}{(2\pi)^4}\frac{1} {\pi^2 - \mu^2 + \i \varepsilon}
\sqrt{\displaystyle\frac{E_{lab}+m}{E_q+m}}\frac{2}{E_q}
\frac{1} {\pi^2 - \mu^2 + \i \varepsilon}
\label{triang5}\\[2mm]
& \times &
\frac{ G_{D^{++}}
(q_0,|{\bf q}|)}{(M_D-2E_q)^2 - 4q_0^2 + \i \varepsilon}
\left( |{\bf q}|\cos\theta +\frac{E_q+m}{E_{lab}+m} |\bp'|\cos^2\theta
\right)\nonumber\\[2mm]
& \times &
f^{M'}_{s_2,s}(p,p_2;\pi ,D').
\nonumber
\end{eqnarray}
Then the invariant amplitude ${\cal M}_{s,M}^{s',M'}$
is computed by using  Eqs.~(\ref{ampF}) - (\ref{spinampl}) and
(\ref{triang4}) - (\ref{triang5}). Within the BS formalism
in numerical calculations of integrals with BS amplitudes one usually
performs a  Wick rotation and all calculations are done in the Euclidean space.
This is possible when singularities in integrands are well located and there are
no poles in the first and third quadrant of $q_0$.
In our case the pion pole is a  ``moving'' one  and accidentally, at
some values of $|{\bf q}|$,  it may
cross the imaginary axis  $q_0$ thus hindering the
standard procedure of a Wick rotation.
In this case one needs to  compute either the contribution of the residue of this pole
or  abandon the Wick rotation and to compute the integral by closing the contour in
the upper half-plane and evaluating the residue in each pole appearing in the integrand.
As  mentioned, in our case in the upper half-plane
there are two poles, one from the nucleon
propagator, another one from the pion propagator.
An analysis of the  contribution of the pion pole has been performed in
ref.~\cite{kolybasov} and it is  found to be relatively small,
and with an accuracy of $\varepsilon_D/\mu$
(where $\varepsilon_D$ is the deuteron binding energy) it may be neglected.
Then the main contribution to  the integrals (\ref{triang4}) and (\ref{triang5})
comes  from the residue at $q_0 = M_D/2 - E_q$, where the exchanged proton is
on mass shell, i.e.  $p_{20} = E_q$, and the
neutron is off mass shell, i.e. $p_{10}=M_D-E_q$.
Here it is worth emphasizing that when calculating
loop and triangle diagrams in quantum field theory,
 after performing integrations
on $q_0$ in one propagator, the second one remains
singular in ${\bf q}$.  These singularities,
known as Landau ghosts \cite{itzixon}, are unphysical and they ought
to be  removed
by   properly choosing  a
subtraction  procedure. In our case these ghosts may appear in
the pion propagator  after the integration
of the nucleon pole  is carried out and the remaining integral
on ${\bf q}$ is extended to infinity.
Within effective meson-nucleon theories  there are no rigorous
regularization schemes to subtract singularities in triangle diagrams.
Usually they are  removed either numerically, for instance by
using cut-off parameters in the integrations,  or by choosing adequate
physical approximations. We proceed now as follows.
At the nucleon pole the pion propagator reads
\begin{equation}
\pi^2-\mu^2 = (E_{lab} - p_{10})^2 -({\bf p}' -{\bf q})^2 -\mu^2=
(E_{lab}+E_q-M_D)^2-({\bf p}' -{\bf q})^2 -\mu^2 .
\label{approximation}
\end{equation}
It is seen that singularities may appear at large values of
$E_q$, which correspond to large $|{\bf q}|$ in the integrals (\ref{triang4}) and
(\ref{triang5});
however  at such  $|{\bf q}|$ the BS waves functions and
 the subprocess $p+p\to \pi + D'$ vanish so that in numerical calculations
this range  of $|{\bf q}|$ should be eliminated.
Following refs.~\cite{nakamura,kolybasov} we introduce
in the laboratory system  the kinetic energies
 of particles on-mass shell, $T_q=E_q-m,\quad T_{lab}=E_{lab}-m$ and due to the fact that
in the process (\ref{reaction}) the momentum of the outgoing proton is
kinematically restricted (since $P_{lab}^2/2m^2 <\, 0.3 $ in the whole range of $\sqrt{S}$)
we neglect in Eq. (\ref{approximation}) terms $\propto P_{lab}^4/4m^4$ and
 $\propto {\bf q}^4/4m^4$, i.e.

\begin{equation}
\pi^2-\mu^2=
(T_{lab}+T_q)^2-({\bf p}' -{\bf q})^2 -\mu^2 \approx -({\bf p}' -{\bf q})^2 -\mu^2.
\label{approximat}
\end{equation}

Plugging Eq.~(\ref{approximat}) into Eqs.
(\ref{triang4}) and (\ref{triang5}) we obtain
\begin{eqnarray}
a_{s_2,s}^{M'} & = &
-\frac{g_{\pi NN}}{(4\pi)^{3/2}} \sqrt{2M_D}
\int\frac{{\bf q}^2 d |{\bf q} | d\cos\theta}{({\bf q}-{\bf p}')^2+\mu^2}
\frac{1}{E_q}\sqrt{\displaystyle\frac{E_{lab}+m}{E_q+m}}
\label{anew}\\[2mm]
& \times &
\left( \Psi_S(q_0,|{\bf q}|)-\Psi_D(q_0,|{\bf q}|) \frac{1}{\sqrt{2}}
\right )
\left( |{\bf q}|\cos\theta +\frac{E_q+m}{E_{lab}+m} |\bp'|\right)
\nonumber\\[2mm]
& \times &
f^{M'}_{s_2,s}(p,p_2;\pi ,D'),\nonumber\\[4mm]
d_{s_2,s}^{M'} & = &
-\frac{6g_{\pi NN}}{(4\pi)^{3/2}} \sqrt{M_D}
 \int\frac{{\bf q}^2 d |{\bf q}|d\cos\theta}{({\bf q}-{\bf p}')^2+\mu^2}
\frac{1}{E_q}\sqrt{\displaystyle\frac{E_{lab}+m}{E_q+m}}
\label{dnew}\\[2mm]
& \times &
\Psi_D(q_0,|{\bf q|})
\left( |{\bf q}|\cos\theta +\frac{E_q+m}{E_{lab}+m} |\bp'|\cos^2\theta\right )
f^{M'}_{s_2,s}(p,p_2;\pi ,D').
\nonumber
\end{eqnarray}
Our ansatz, Eq.~(\ref{approximat}),  is  very similar to that one adopted
in ref.~\cite{nakamura}, however, there is a difference
in the treatment of the kinetic energy of the off-mass shell neutron.
In  ref.~\cite{nakamura} the neutron kinetic energy has been taken to be
identical to the one of the exchanged proton. This  leads to a violation of the energy
conservation of the deuteron in the laboratory frame.
However,  at low and moderate values of $P_{lab}$
numerically both approximations provide nearly the same results.
Note  that in the extreme non-relativistic case,
${\bf q}^2 \ll m^2$,
Eq.~(\ref{approximat}) yields  $\pi^2-\mu^2\approx -(2mT_{lab}+\mu^2)$,
i.e. exactly the pion propagator  used in ref.~\cite{kolybasov}.

Finally, after an explicit evaluation of the spin
matrix elements in Eq.~(\ref{simple})
(see also, Appendix \ref{sec:app2}) the  amplitude corresponding to
the diagram Fig.~1b may be cast in the form
\begin{eqnarray}
T_{s,M}^{s',M'} & = &
\sum_{s_2}
\left\{
\delta_{M0}
\left( a_{s_2,s}^{M'}+d_{s_2,s}^{M'} \right)
\left( \delta_{s_2\frac{1}{2}} \delta_{s'-\frac{1}{2}}
-\delta_{s_2-\frac{1}{2}}\delta_{s'\frac{1}{2}}
\right) \right.
\nonumber\\[2mm]
& + &
\left.
a_{s_2,s}^{M'}
\left(
-\sqrt{2}\delta_{M1}\delta_{s_2\frac{1}{2}} \delta_{s'\frac{1}{2}}
+\sqrt{2}\delta_{M-1}\delta_{s_2-\frac{1}{2}} \delta_{s'-\frac{1}{2}}
\right)
\right\} ,
\label{tot}
\end{eqnarray}
with $ a_{s_2,s}^{M'}$ and  $ d_{s_2,s}^{M'}$ given by Eqs.~(\ref{anew}) and
(\ref{dnew}), respectively.

For explicit calculations of the amplitude (\ref{tot}) the amplitude
$f^{M'}_{s_2,s}(p,p_2;\pi ,D')$ of the elementary process $pp\to \pi D$ is
needed at an unphysical value of the mass of the virtual  pion, and so there
is an ambiguity related to the value to be used. Since  the amplitude depends on two
invariant variables, and the Mandelstam $u$ is common for both the
process (\ref{reaction}) and the subprocess  $pp\to \pi D$, we choose the
independent variables for $f$ to be $u$ and $s_{12}=(p+p_2)^2$. Then assuming
that the amplitude does not vary as a function of the pion mass, one
may reconstruct the amplitude by using the experimental data for the real
process  $pp\to \pi D$ at given $s_{12}$ and $u$.
In our calculation we use the partial amplitudes
$f^{M'}_{s_2,s}(s_{12},u)$ from the combined analysis of Arndt et al.\cite{arndt},
which are  available via telnet in an interactive regime
(see references in \cite{arndt}).  In our calculations we need the
partial amplitudes in the laboratory frame where the vectors $\bp $ and $\bp_2$
are not collinear, while the experimental amplitudes are given in the
center of mass of colliding particles. Consequently, a Lorenz boost of the
amplitudes from the center of mass to the laboratory system is needed. As a result
two additional rotations of the amplitude
(the Wick helicity rotation \cite{bourrely}
to adjust the helicity amplitudes in two systems, and a Wigner rotation from
$z$ axis parallel to $(\bp + \bp_2)$ to the $z'$ axis along the vector
$\bp'$) have to be employed (see Appendix \ref{sec:app3}).

The contribution to the invariant amplitude $\CM$ from the
diagram Fig.~1c is computed in the frame where the outgoing deuteron
is at rest and it is found to be related  to
the contribution of the diagram 1b via
\begin{equation}
\widetilde T_{s,M}^{s',M'} = T_{-s',-M'}^{-s,-M}.
\label{time}
\end{equation}
It is seen  that the sum of the two diagrams 1b and 1c
ensures the total  amplitude ${\CM}$ to be  time reversal invariant.

\section{Results}

In Fig.~2 results of numerical calculation of the differential cross
section are presented. The dashed line is the contribution of the
one-nucleon exchange mechanism \cite{ourphysrev}. It is seen that
within the impulse approximation a satisfactorily description
of data cannot be achieved. The contribution
of one-meson exchange diagrams (dotted line)  is significant
in the region $P_{lab} \sim 0.2 \,-\, 0.4$ GeV/c (corresponding to the range
of the initial momentum    $p\sim 0.66\, - \, 1.85$ GeV/c),  which is
the region (enlarged by ''Fermi motion'' effects) where the
experimental data give evidence
for  $\Delta$ excitations in  the amplitude $p+p\to \pi^+ +D$
\cite{arndt}. Beyond this region the amplitude
$p+p\to \pi^+ +D$ vanishes rather rapidly as also the
triangle amplitude does. The total cross section
is represented by solid lines, where the line labeled by (a) is the result
of calculations with taking into account the full BS solution
(including $P$ waves), the  label  (b) depicts the BS result with
only positive-energy waves. It may be seen that an account for
$\Delta$ excitations (through the $p+p\to \pi^+ +D$ amplitudes) essentially
improves the description of the experimental data.

Contrary to this, the polarization observables are less sensitive
to the contribution of the triangle diagrams. As seen in Figs.~3 and 4 the
agreement with data of tensor analyzing power $T_{20}$ and polarization transfer
$\kappa_{D\to p}$ is not improved.
Fig.~5 demonstrates that also the polarization transfer $\kappa_{p\to p}$ is
essentially unaffected by the  triangular diagrams.
To understand this insensitivity one has to
recall that within the one-nucleon exchange approximation the corresponding expressions for the
polarization observables (but not for the cross sections and with
positive BS waves only) are analytically identical for all the
backward $pD$ processes, elastic, quasi-elastic and break-up
reactions~\cite{ourphysrev}. This is due to a factorization of contributions
from upper and lower vertices in diagrams like the one depicted in Fig.~1a.
The triangle diagram, Eqs.~(\ref{anew}) and (\ref{dnew}), is
implicitly proportional to the second power of the deuteron wave function
(the amplitude $f_{pp\to \pi^+ D}$ itself is proportional to the
first power of the wave function of the outgoing deuteron), so that
the contribution to the cross section is $\propto \Psi_D^4$ as is the
cross section in the one-nucleon exchange approximation. However, our
numerical results for polarization observables, Figs.~3 - 5, and a comparison
with those obtained within one-nucleon exchange approximation \cite{ourphysrev},
persuades us that the amplitude for triangle diagrams may be
also approximately presented as a factorization between the upper vertex
(the amplitude $f_{pp\to \pi^+ D}$ in Fig.~1b)
and the lower vertex, characterizing
the structure of the target deuteron. Consequently, potentially large
effects in the cross section cancel in the ratios, which define
the polarization observables.
Moreover, it seems that canonical approaches,
treating the  deuteron solely by nucleon
degrees of freedom, should be supplemented by different ``exotics''
(e.g. $\Delta$ - $\Delta$ components). More experimental data on polarization
characteristics of backward $pD$ reactions will provide more complete information
about the deuteron structure and reaction mechanisms.

\section{Summary}

In summary, we present an explicit analysis of pion-exchange  effects
in  elastic backward scattering of protons off deuterons  within the Bethe-Salpeter
formalism with realistic interaction kernel.
The total amplitude of the process is
presented as a sum of contributions of the one-nucleon
exchange mechanism and the triangle diagrams with virtual
$\Delta$ excitations. The four partial  spin amplitudes of the process
have been computed explicitly within the Bethe Salpeter approach.
Effects of relativistic Fermi motion and Lorenz transformations
of the amplitudes have been taken into account in a fully covariant way.
Numerical estimates of effects  of pion exchanges
in the cross section and polarization observable,
e.g. tensor analyzing power and polarization transfers,
at kinematical conditions of operating \cite{proposal} and
forthcoming experiments \cite{cosy},
 are performed. It is found that the one-pion exchange mechanism
plays a crucial role in describing the spin averaged cross section, while
for the considered polarization observables the role of triangle
diagrams is less important.

It is shown that the one-nucleon and one-pion exchange exchange mechanisms
are not the  predominant ones in describing various polarization
observables  and future experiments
mainly will highlight effects beyond these mechanisms.

\begin{acknowledge}

We thank R. Arndt and I. Strakovsky for useful discussions
and explanations of how to use the SAID program
in an interactive regime to obtain their partial amplitudes for the
$pp\to \pi D$ reaction. We especially thank R. Arndt which provide us
with Fortran codes to compute the helicity amplitudes $f^{M'}_{s_2,s}$.
Useful discussions with
A.Yu. Umnikov, Yu. Kalinovsky, L. Naumann and F. Santos
are gratefully acknowledged.
Two of the authors (L.P.K. and S.S.S.) would like
to thank for the warm hospitality in the Research Center Rossendorf.
This work is supported
within the Heisenberg-Landau JINR-FRG collaboration project, and by
BMBF  06 DR 829/1 and RFBR No.95-15-96123.

\end{acknowledge}

\appendix

\section{The Bethe Salpeter Amplitudes}
\label{sec:app1}

The Bethe Salpeter amplitudes in the laboratory system are of the
form \cite{quad}
\begin{eqnarray}
\Phi_M^{S^{++}}(p_1,p_2) & =&
{\cal N}(\hat k_1+m )\frac{1+\gamma_0}{2}\hat\xi_M(\hat k_2-m)
\phi_S (p_0,|{\bf p}|), \label{psis}\\[2mm]
\Phi_M^{D^{++}}(p_1,p_2) & = &
-\frac{{\cal N}}{\sqrt{2}}
(\hat k_1+m )\frac{1+\gamma_0}{2}
\left (
\hat\xi_M +\frac{3}{2|{\bf p}|^2} (\hat k_1-\hat k_2)(p\xi_M)\right )
(\hat k_2-m)
\phi_D (p_0,|{\bf p}|),\nonumber
\end{eqnarray}
where
$k_{1,2}$ are on-mass shell four-vectors related  to the
off-mass shell vectors $p_{1,2}$ as follows
\begin{equation}
k_1=(E,\bp),\quad k_2=(E,-\bp),\quad p_1=(p_{10},\bp),\quad
p_2=(p_{20},-\bp),\quad E=\sqrt{\bp^2+m^2},
\nonumber
\end{equation}
and $\phi_{S,D} (p_0,|{\bf p}|)$ are the partial scalar amplitudes, related to
the corresponding partial vertices as
\begin{equation}
\phi_{S,D} (p_0,|{\bf p}|)=
\displaystyle\frac{G_{S,D} (p_0,|{\bf p}|)}{\left(\displaystyle\frac{M_D}{2}-E\right)^2-p_0^2}.
\nonumber
\end{equation}
In Eq.~(\ref{psis}) the normalization factor is
${\cal N}=\displaystyle\frac{1}{\sqrt{8\pi}}\displaystyle\frac{1}{2E(E+m)}$.

\section{The  spinor matrix elements}
\label{sec:app2}

Here we present some explicit formulae which may be useful in computing
the spinor matrix elements in Eq.~(\ref{triang3}) and in
deriving Eq.~(\ref{tot}):
\begin{eqnarray}
&&
\widetilde\chi_s = -\i \sigma_y\,\chi_s;
\quad -\i \sigma_y = \displaystyle\frac{\sigma_+ + \sigma_-}{\sqrt{2}};
\quad \sigma_{+(-)}=\displaystyle\frac{-(+)\sigma_x-i\sigma_y}{\sqrt{2}},
\\[2mm]
&&
(\bsigma\,{\bn})(\bsigma\,\bxi_M)=(\bn\,\bxi_M) +
\i \left( \bsigma\,[\bn\times\bxi_M]\right );
\nonumber\\[2mm]
&&
-\i \left( \bsigma\,[\bn\times\bxi_M]\right )\,
\i \sigma_y=(\bxi_M)_x
-\i (\bxi_M)_y\sigma_z
=
\left \{
\begin{array}{cc}
-\sqrt{2}\hat P_+ & \quad M=+1\\
0                 &  \quad M=0\\
\sqrt{2}\hat P_-  & \quad M=-1
\end{array}
\right . ,
\end{eqnarray}
where $\hat P_{\pm}$ is the spin-$1/2$ projection operator on states with
positive (negative) spin projections, $\hat P_{\pm} = (1\pm\sigma_z)/2$.

\section{Helicity Wick rotation  \lowercase{\protect\cite{bourrely}} }
\label{sec:app3}
By definition, a state with given momentum $\bp$ and helicity $\lambda$
in a frame of reference $\CO$ is that obtained by
a Lorenz transformation of a state with given spin projection
$s_z$ from the rest system
$\CO_{rest}$ to $\CO$, i.e.:

\begin{equation}
|\bp;\lambda\rangle\equiv | \stackrel{0}{p},s,s_z\rangle_\CO
\end{equation}
where $\stackrel{0}{p}=(m,0,0,0)$. As usually, a Lorenz transformation $h[\bp]$
is presented by a sequence of two operations:
a boost along the $z$ axis, $l_z(v)$, where $v$ is
the speed of the state in $\CO$, and a rotation from $z$ direction to the
direction of $\bp$, i.e.
$\CO=r^{-1}(\phi,\theta,0)l_z^{-1}(v)\CO_{rest}$.

Let us suppose now
that one has a state $|\bp;\lambda\rangle$ given in the frame $\CO$ and
one wishes to know how it reads in another frame  $\CO'$ obtained
by a Lorenz transformation
$l$ on $\CO$
\begin{equation}
|\bp;\lambda\rangle_{\CO'} = U(l^{-1})|\bp;\lambda\rangle.
\end{equation}
From the definition of the helicity states one has
\begin{equation}
U(l^{-1})|\bp;\lambda\rangle =
U(l^{-1}) U(h[(\bp)])| \stackrel{0}{p};\lambda\rangle ,
\label{b3}
\end{equation}
where $h[\bp]$ is the corresponding Lorenz transformation $\stackrel{0}{p}\to p$.
Then multiplying Eq.~(\ref{b3}) by unity,
$U(h[(\bp')])U^{-1}[h[(\bp')]=1$, where $h[\bp']$
is the helicity transformation that would define a state
$|\bp';\lambda\rangle =U(h[(\bp'])| \stackrel{0}{p},\lambda\rangle $ with
$\bp'$ being the same vector as obtained by transforming $\bp$ from
$\CO$ to $\CO'$,
one obtains:
\begin{equation}
U(l^{-1})|\bp;\lambda\rangle =U(h[(\bp']){\cal R}| \stackrel{0}{p},\lambda\rangle ,
\end{equation}
where
${\cal R}=U^{-1}[h[(\bp')]U(l^{-1})U[h(\bp)]$ is a sequence of transformations
$ \stackrel{0}{p}\to p \to p' \to  \stackrel{0}{p}$, i.e. nothing but a rotation.
Then
\begin{equation}
|\bp,\lambda\rangle_{\CO'} = D^{(s)}_{\lambda\lambda'}(\omega)
|\bp',\lambda'\rangle,
\end{equation}
where $\omega$ is a set of Euler angles describing the rotation.
In case when the Lorenz transformation is a simple boost along the $z$ direction
with the speed $\beta$ then $\omega$ is just an angle, describing a rotation
about the $Y$ axis,
\begin{eqnarray}
\cos\omega=\cos\theta'\cos\theta+ \gamma\sin\theta'\sin\theta,
\end{eqnarray}
with $\gamma=1/\sqrt{1-\beta^2}$ and $\theta, \theta'$ are the polar angles of $\bp$
in the systems $\CO$ and $\CO'$, respectively.
This is known as Wick helicity rotation,
contrary to Wigner's canonical spin rotation.
In our case the relevant $z$ axis is the one along the  $(\bp_2+\bp)$ direction.
Then, obtaining the helicity amplitudes in the laboratory frame we need an additional
rotation to change from the helicity basis to the spin projections.

\newpage

\newpage
\setcounter{section}{0}
\begin{figure}[hb]
\vskip 1cm
\epsfxsize 1.6in
\hspace*{-1.5cm}\epsfbox{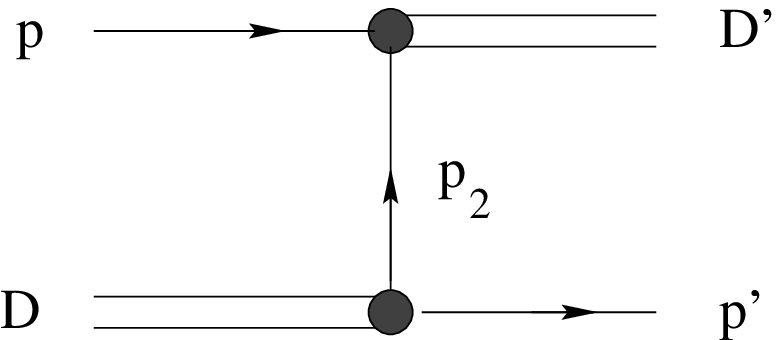}

\vspace*{-2.5cm}
\hspace*{2cm}\epsfxsize 2.9in
\epsfbox{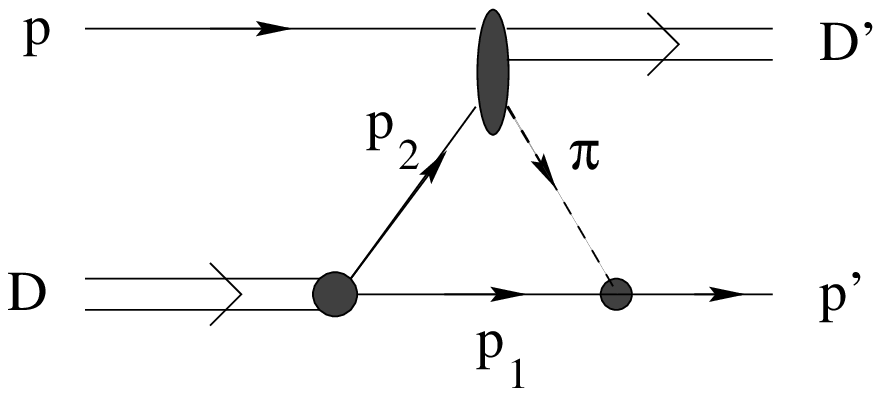}

\vspace*{-2.5cm}
\hspace*{10cm}\epsfxsize 2.0in
\epsfbox{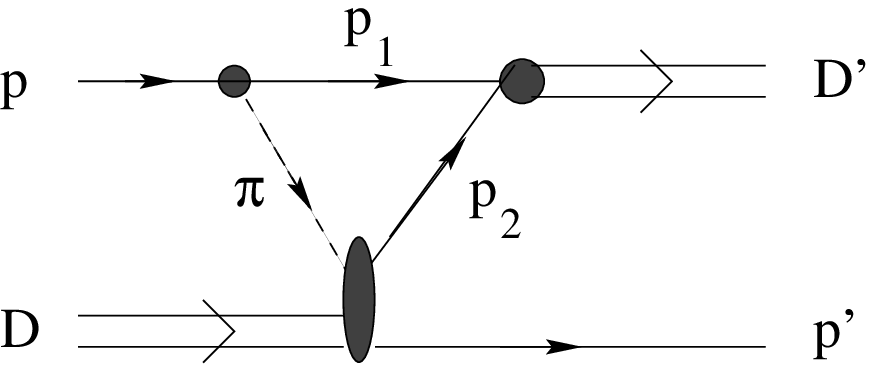}

\vspace*{1.5cm}
\hspace*{-.1cm} (a) \hspace*{6cm} (b) \hspace*{4.5cm} (c)
\label{diag}

\vspace*{5mm}
\caption{
The one-nucleon (a) and the   one-pion  exchange   graphs (b) and (c)
for the reaction
$p\, +\, D\, =\, p'(\theta^*=180^o)\,+\,D'$.}
\end{figure}

\begin{figure}[hb]
\epsfxsize 4in
\begin{center}
\epsfbox{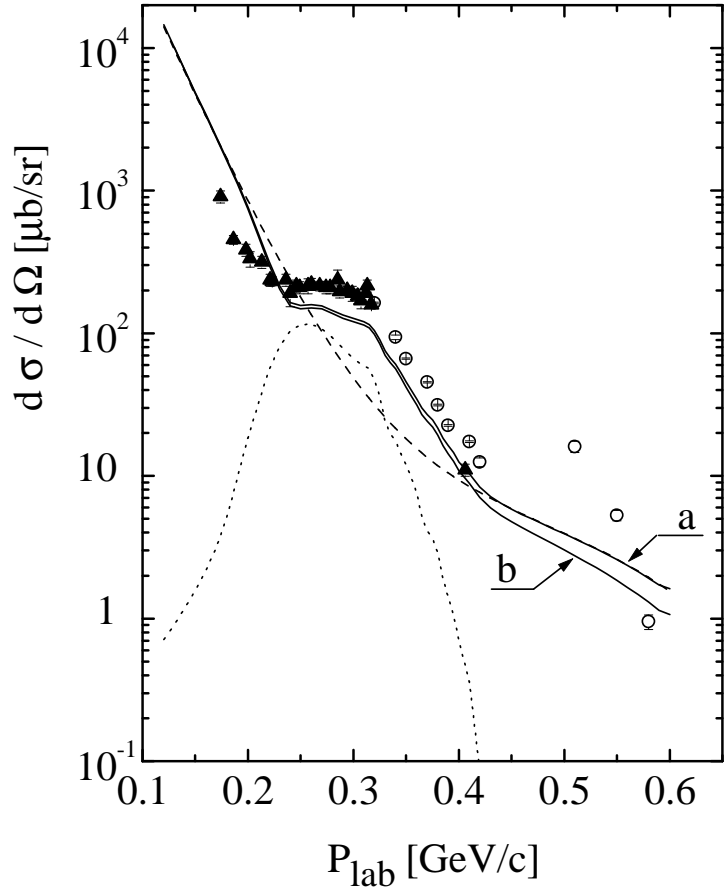}
\label{crossfig}
\end{center}

\caption{
The spin averaged differential cross section
$\displaystyle\frac{d\sigma}{d\Omega}$
for the elastic $pD$ backward scattering
in the center of mass system as a function
of the momentum of the detected proton in the laboratory system.
Dashed line:   the contribution of the
one-nucleon exchange mechanism; dotted line:
results of calculations of the triangle diagram;
solid line: the full BS calculations  with $P$ waves (a), or
without $P$ waves (b);
The experimental data are from \protect\cite{dpelasexp,nakamura}.}
\end{figure}

\newpage

\begin{figure}[ht]
\epsfxsize 4in
\begin{center}
\epsfbox{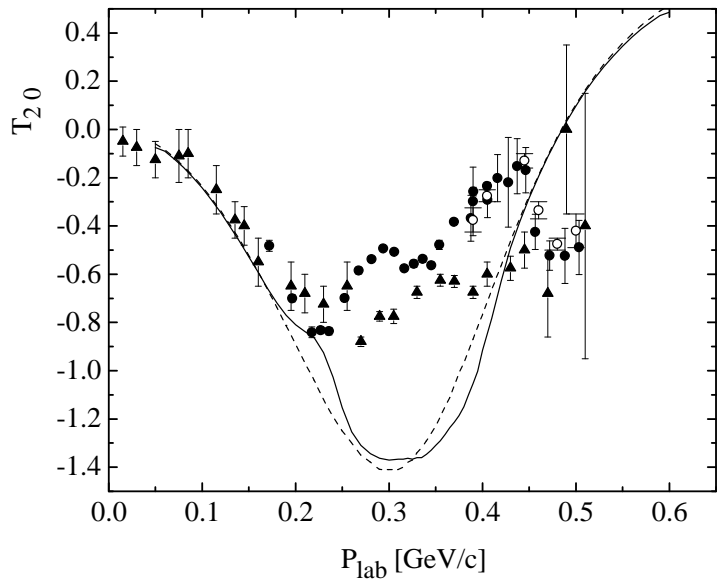}
\label{t20fig}
\end{center}
\caption{
The deuteron tensor analyzing power $T_{20}$
for the elastic $pD$  scattering.
Dashed line: the contribution of the one-nucleon exchange mechanism;
solid line: the
result of computation within
the BS approach, including contributions from the
triangle diagram.
Experimental data: circles - $T_{20}$ for the elastic
backward scattering, ref.~\protect\cite{dpelastic,punj95,azh97},
triangles -  $T_{20}$ measured in the deuteron
break up reactions \protect\cite{experiment}.}
\end{figure}

\newpage

\begin{figure}[ht]
\epsfxsize 4in
\begin{center}
\epsfbox{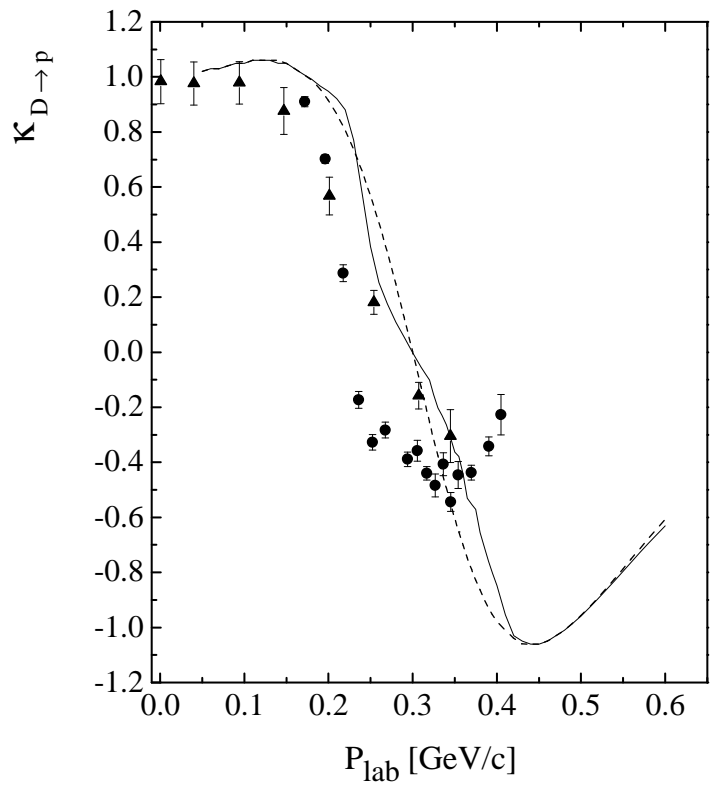}
\label{kappafig}
\end{center}
\caption{
The deuteron to proton polarization transfer $\kappa_{D\to p}$
for the elastic $pD$ backward scattering.
Notation as in Fig.~3.}
\end{figure}
\newpage

\begin{figure}[ht]
\epsfxsize 4in
\begin{center}
\epsfbox{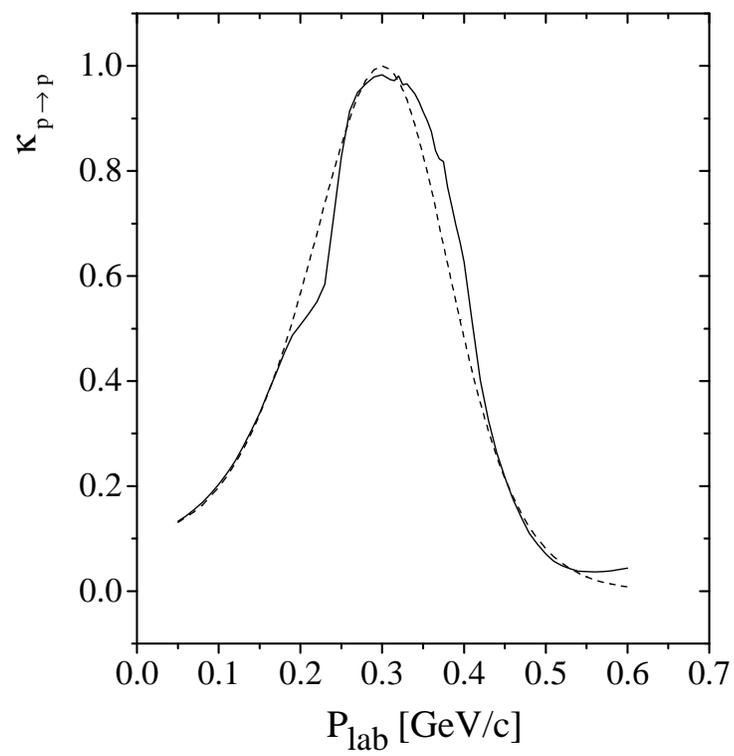}
\label{drkappafig}
\end{center}
\caption{
The vector-vector polarization transfer coefficient
from the initial proton to the final proton.
Dashed line: contribution of the one-nucleon exchange mechanism;
solid line: the full BS results.}
\end{figure}
\end{document}